\documentclass{iopart}
\usepackage{graphicx}
\begin{document}
\title[Self-similar analytical solution of the critical fluctuations problem for BEC]{Self-similar analytical solution of the critical fluctuations problem for the Bose-Einstein condensation in an ideal gas}
\author{Vitaly V Kocharovsky$^{1,2}$ and Vladimir V Kocharovsky$^2$}
\address{$^{1}$Department of Physics, Texas A\&M University, College Station, TX 77843-4242}
\address{$^{2}$Institute of Applied Physics, Russian Academy of Science,
603950 Nizhny Novgorod, Russia}
\ead{vkochar@physics.tamu.edu}
\begin{abstract}
Paper is published in J. Phys. A: Math. Theor. 43 (2010) 225001, doi:10.1088/1751-8113/43/22/225001

  Exact analytical solution for the universal probability distribution of the order parameter fluctuations as well as for the universal statistical and thermodynamic functions of an ideal gas in the whole critical region of Bose-Einstein condensation is obtained. A universal constraint nonlinearity is found that is responsible for all nontrivial critical phenomena of the BEC phase transition. Simple analytical approximations, which describe the universal structure of the critical region in terms of confluent hypergeometric or parabolic cylinder functions, as well as asymptotics of the exact solution are derived. The results for the order parameter, all higher-order moments of BEC fluctuations, and thermodynamic quantities, including specific heat, perfectly match the known asymptotics outside critical region as well as the phenomenological renormalization-group ansatz with known critical exponents in the close vicinity of the critical point. Thus, a full analytical solution to a long-standing problem of finding a universal structure of the $\lambda$ point for BEC in an ideal gas is found.
\end{abstract}
\submitto{Journal of Physics A: Mathematical and Theoretical}

\section{Critical fluctuations and universal structure of the $\lambda$ point}
Fluctuations in finite-size systems and their microscopic theory have attracted great interest in recent years due to numerous modern experiments on Bose-Einstein condensation (BEC) in the traps which contain usually a finite number of atoms $N$ ranging from a few hundred to a few million (see, e.g., [1-3] and references therein). Solution of that problem is equivalent to a solution of a long-standing problem of finding the microscopic theory of critical phenomena in the second-order phase transitions, including Bose-Einstein condensation, which is still missing despite some success of the phenomenological renormalization-group theory. The problem of finding the universal functions of the statistical and thermodynamic quantities in the critical region of BEC is not solved yet even for an ideal Bose gas although the problem itself is almost a century old and BEC phase transition in an ideal gas is considered to be a basic chapter of statistical physics (see [4-6] or any other textbook). A nice recent paper \cite{Kleinert2007} by Glaum, Kleinert, and Pelster, and the comprehensive textbook by Kleinert \cite{Kleinert} clearly present a modern status of this problem, including the problem of the structure of heat capacity near the $\lambda$ point as well as the first-quantized path-integral imaginary-time formalism. They conclude that for the solution of this problem "analytical expression within the canonical ensemble could not be found, so we must be content with the numerical results". 

     Previous attempts to solve the problem, for example the ones based on a grand-canonical-ensemble approximation in the thermodynamic limit \cite{LLV,Pathria} or a phenomenological renormalization-group approach [4, 9-14], did not resolve the universal fine structure of the critical region or did not provide the analytical formulas for the universal functions in the whole critical region. The renormalization-group approach uses Monte Carlo or other simulations for relatively small finite-size systems and fits these simulation data to some finite-size scaling ansatz which typically involves only one or a few first derivatives of the universal functions at the critical point and is not valid in the whole critical region (for the examples related to BEC, see [14-21]). Here we present a full analytical solution to this problem for an ideal gas, including the exact analytical formulas for the universal functions of the statistical and thermodynamic quantities in the whole critical region. In particular, we find a full analytical solution to the long-standing problem of resolving a fine universal structure of heat capacity near the $\lambda$ point for BEC in an ideal gas.

    Contrary to a case of a macroscopic system in the thermodynamic limit or a case of a few-body system, a general case of a mesoscopic number of atoms is especially difficult for it requires a solution that explicitly depends on the number of atoms. The standard in statistical physics methods, including the grand-canonical-ensemble method and Beliaev-Popov method of quantum field theory \cite{LL,Ziff,Shi}, are not appropriate for the analysis of critical fluctuations (see, e.g., \cite{PitString,Koch06,Kleinert2007,Ziff,Sinner,Balazs1998}). It has been demonstrated already by an analysis of a well-known grand-canonical catastrophe of the BEC fluctuations \cite{PitString,Koch06,Ziff}. In fact, to solve the problem of critical fluctuations in BEC phase transition one has to find an explicit solution to a statistical problem of BEC for a finite number of atoms in the trap in a canonical ensemble. A full physical picture of the statistics of BEC has been absent not only in a general case of an interacting gas, but even in the case of an ideal gas (see, e.g., [2, 7, 21-34] and references therein). A formula for the anomalously large variance of the ground-state occupation found both for an ideal gas (see \cite{Ziff} and references therein) and for a weakly interacting gas [35-38], is valid only far enough from the critical point, where fluctuations of the order parameter are already relatively small. The same is relevant also to a known result on the analytical formula for all higher-order cumulants and moments of BEC fluctuations, which demonstrates that BEC fluctuations are essentially non-Gaussian even in the thermodynamic limit \cite{KKS-PRL,KKS-PRA}. Also, the probability distribution of the order parameter or the logarithm of that distribution, i.e., a Landau function \cite{Goldenfeld}, for the BEC in an ideal gas in the canonical ensemble has been discussed in literature \cite{Koch06,ww1997,Holthaus1997,Balazs1998,KKS-PRL,KKS-PRA,Wilkens2000,Baym2001,Sinner}. However, its full analytical picture in the whole critical region has not been found. The renormalization-group ansatz for the finite-size scaling variable for the interacting and ideal gases has been used [15-21, 31], but the universal structure of BEC statistics in an ideal gas was found only recently \cite{KKD-RQE}.
     
    Due to space limitation and in order to present the solution of the critical-fluctuations problem in the most transparent form, we consider here only the case of an ideal gas. Moreover, in the present paper we focus only on the critical fluctuations in macroscopic systems in the thermodynamic limit. Analytical theory of mesoscopic BEC fluctuations with detailed description of all discrete and finite-size effects will be discussed elsewhere. 
    
    The problem of critical fluctuations in a weakly interacting gas can be solved on the basis of the same method and in terms of the same functions as we use here for an ideal gas since the constraint nonlinearity is the basic reason for critical behaviour in any second-order phase transitions. The solution for a weakly interacting gas will be presented in a separate paper. Here we make only a few short remarks concerning the case of an interacting gas. First, a remarkable fact that a simple ideal gas system demonstrates BEC phase transition even without interparticle interaction allows us to find the analytical solution and the universal constraint-cutoff origin of the nonanalyticity and critical phenomena in all other interacting systems demonstrating second-order phase transition. Second, it is known that the particular features of the interparticle interactions are not very important for the universal properties of a phase transition and, for example, do not affect a universality class of a phase transition. Nevertheless, we have to take into account a deformation of the statistical distribution due to a feedback of the order parameter on the quasiparticle energy spectrum and interparticle correlations. Third, to do so in a nonperturbative-in-fluctuations way we use a theorem on the nonpolynomial averages in statistical physics and an appropriate diagram technique proposed in \cite{KochLasPhys2007,KochJMO2007}. 
    
\section{Canonical-ensemble quasiparticles}
The most crucial issue for the solution of the BEC phase transition problem is the exact accounting of the particle-number constraint ${\hat n}_{0} + \sum_{\vec{k}\neq 0}{{\hat n}_{\vec{k}}} = {\hat N} = const$ as an operator equation which is responsible for the very BEC phenomenon and is equivalent to an infinite set of the c-number constraints. It cannot be replaced by just one condition for the mean values, $N = {\bar n}_{0} + \bar n$, used in the grand-canonical-ensemble approach to specify an artificially introduced chemical potential $\mu$. Here, in the canonical ensemble, all averages are understood as taken over an equilibrium canonical-ensemble density matrix ${\hat \rho} = exp(-{\hat H}/T)/Tr[exp(-{\hat H}/T)]$, where ${\hat H}$ is the hamiltonian operator and $T$ the temperature measured in the energy units. An operator ${\hat N} = N {\hat I}$ is defined as a product of a fixed number of atoms in the trap $N$ and a unity operator ${\hat I}$. Obviously, its mean value ${\bar N} = N$ is nonzero, but all central moments of the operator ${\hat N}$ are equal to zero. Following commonly accepted notations, we denote ${\hat N}$ simply by $N$ omitting the unity operator ${\hat I}$ (or a hat above ${\hat N}$) in all operator equations which involve ${\hat N}$. Contrary to ${\hat N}$, an occupation operator ${\hat n}_{\vec{k}}= \hat{a}_{\vec{k}}^{\dagger} \hat{a}_{\vec{k}}$ for the $\vec{k}$ state of an atom in the trap (${\hat a}_{\vec{k}}$ and ${\hat a}_{\vec{k}}^{\dagger}$ are annihilation and creation operators for an atom in the $\vec{k}$ state, respectively) and an operator $\hat n = \sum_{\vec{k}\neq 0}{{\hat n}_{\vec{k}}}$ of a total occupation of all excited atomic states are nontrivial fluctuating quantities. The problem is to find their statistics and, first of all, the statistics of the condensate occupation ${\hat n_0} = N - {\hat n}$. The probability $\rho_n$ to find $n$ atoms in the excited states, that is in the noncondensate, is equal to the probability $\rho^{(cond)}_{n_0 =N-n}$ to find $n_0 = N-n$ atoms in the condensate. It is more convenient to work with $\rho_n$ since the total noncondensate occupation $n$ is a sum of many, almost independent occupations $n_{\vec{k}}$. 

   We solve for the constraint from the very beginning by a proper reduction of the many-body Hilbert space following our general approach \cite{Koch06,KKS-PRL,KKS-PRA,KochLasPhys2007,KochJMO2007}. In the present case of an ideal gas in the canonical ensemble, we have to consider as independent only noncondensate Fock states $\left| \left\{ n_{\vec{k}}, \vec{k} \neq 0\right\} \right\rangle$  which uniquely specify the ground-state Fock state $\left| n_0 = N - \sum_{\vec{k}\neq 0}{n_{\vec{k}}}\right\rangle$. Physical processes in the trap proceed through creation or destruction of the excitations in the system of a fixed number of atoms $N$, not through creation or destruction of atoms. These excitations are the canonical-ensemble quasiparticles which correspond to the transitions between ground ($\vec{k} =0$) and excited ($\vec{k} \neq 0$) atomic states and are described by the particle-number concerving creation and annihilation operators \cite{ga,g} 
\begin{equation}
\hat{\beta}_{\vec{k}}^{\dagger} =
\hat{a}_{\vec{k}}^{\dagger} \hat{\beta}_0 , \qquad \hat{\beta}_{\vec{k}} =
\hat{\beta}_0^{\dagger}\hat{a}_{\vec{k}} , \qquad \hat{\beta}_0 = ( 1 + \hat{n}_0
)^{-1/2} \hat{a}_0 .
\label{betaoperators}
\end{equation}
They obey the Bose canonical commutation relations everywhere, except the subspace $n_{0}=0$ which has negligible contribution for all interesting temperatures, including even temperatures much higher than a critical temperature $T_c$ of BEC. For the ${\vec k}$ state, the number of canonical-ensemble quasiparticles coincides with the atomic occupation: $\hat{n}_{\vec{k}} = \hat{a}_{\vec{k}}^{\dagger} \hat{a}_{\vec{k}} = \hat{\beta}_{\vec{k}}^{\dagger}\hat{\beta}_{\vec{k}}$. 

To be specific, we consider here an equilibrium ideal gas of $N$ Bose atoms trapped in a cubic box with a volume $V=L^3$ and periodic boundary conditions for which the discrete one-particle energies are equal to $\epsilon_{\vec{k}} = \hbar^2 k^{2}/(2m)$, where $m$ is a mass of an atom and $\vec{k} = 2\pi \vec{q}/L$ a wave vector with the components $k_i = 2\pi q_{i}/L, q_i = 0, \pm 1, \pm 2, \ldots$. This mesoscopic system is described by a Hamiltonian ${\hat H} = \sum^{\infty}_{k=0}{\epsilon_{\vec{k}} {\hat n}_{\vec{k}}}$. In a well-known model of an ideal gas \cite{LLV,Pathria} the temperature is well defined due to an infinitesimally weak interaction of the gas with an external reservoir of temperature $T$ that allows for an equilibration in a long enough time, but does not alter the properties of critical phenomena. 
  
\section{Many-body Hilbert space cutoff and constraint nonlinearity}
As we discussed in \cite{KochLasPhys2007,KochJMO2007}, the only reason for the BEC of atoms on the ground state $\vec{k} = 0$ is conservation of the total number of Bose particles in the trap, $N = {\hat n}_0 + \hat n$. Hence, the occupation operators ${\hat n}_{\vec{k}}$ are not independent and the many-body Hilbert space is strongly constrained. A more convenient equivalent formulation of the problem can be given if one introduces a constraint nonlinearity in the dynamics and statistics, even for the ideal gas, on the basis of the particle-number constraint. Namely, the Fock space of canonical-ensemble quasiparticles, i.e., noncondensate excitations, should be further cut off by the boundary $\sum_{\vec{k}\neq 0}{n_{\vec{k}}} \leq N$ since the number of quasiparticles cannot be larger than the number of atoms in the trap. That cutoff is equivalent to an introduction of a step function $\theta (N-{\hat n})$, i.e., 1 if $n \leq N$ or 0 if $n > N$, in all operator equations and under all trace operations. The $\theta (N-{\hat n})$ factor is the constraint nonlinearity that allows us to consider the noncondensate many-body Fock space formally as an unconstrained one. 
     
    On this basis we immediately obtain the exact solution for the probability to find $n$ atoms in the noncondensate, i.e., $n_0 =N-n$ atoms in the condensate,  
\begin{equation}
\rho_n =\frac{\rho_{n}^{(\infty)}\theta (N-n)}{\sum_{n=0}^{N}{\rho^{(\infty)}_{n}}}.
\label{rhocut}
\end{equation}
It is merely a $\theta (N-{\hat n})$ cutoff of the unconstrained probability distribution for an infinite interval of occupations $n \in \left[ 0, \infty \right)$, found in \cite{KKS-PRL,KKS-PRA} for an arbitrary trap,
\begin{equation}   
\rho^{(\infty)}_{n} =\frac{1}{2\pi}\int_{-\pi}^{\pi}{e^{-iun}\Theta^{(\infty)}(u)du},\quad \Theta^{(\infty)}(u)= \prod_{\vec{k}\neq 0}{\frac{e^{\epsilon_{\vec{k}}/T} - 1}{e^{\epsilon_{\vec{k}}/T} - e^{iu}}}. 
\label{rhoinfinity}
\end{equation}
In other words, for an ideal Bose gas in the canonical ensemble the effective fluctuation Hamiltonian, i.e. the Landau function \cite{Goldenfeld,Sinner}, $-\ln \rho_n$, has an infinite potential wall at $n=N$ and is highly asymmetric due to the constraint nonlinearity. Thus, it remains to fulfill only a straightforward calculation of the moments and cumulants of the cutoff probability distribution, given in (\ref{rhocut}) and depicted as a curve OAN in figure 1a, in order to find the actual BEC statistics in the mesoscopic system for all numbers of atoms and temperatures, including the whole critical region. The probability distribution $\rho_n$ is related to its characteristic function $\Theta(u)$ via Fourier transformation:
\begin{equation}
\rho_n =\frac{1}{2\pi}\int_{-\pi}^{\pi} {e^{-iun}\Theta (u)du},\quad \Theta (u)=\sum^{N}_{n=0}{e^{iun}\rho_{n}}.
\label{rhoFourier}
\end{equation}
The cumulants $\kappa_m$ and the generating cumulants $\tilde{\kappa}_m$ are determined by the Taylor expansion of the logarithm of the characteristic function, $\ln \Theta (u) = \sum_{m=1}^{\infty}{\kappa_m \frac{(iu)^{m}}{m!}} = \sum_{m=1}^{\infty}{\tilde{\kappa}_m \frac{(e^{iu} - 1)^{m}}{m!}}.$ They are related to the central moments $\mu_m =\left\langle (n - {\bar n})^m \right\rangle$ by means of the binomial coefficients and Stirling numbers \cite{KKS-PRL,KKS-PRA,a}. The generating cumulants of $\rho_{n}^{(\infty)}$ for the ideal gas in arbitrary trap were found in \cite{KKS-PRL,KKS-PRA}, 
$\tilde{\kappa}_{m}^{(\infty)} =(m-1)!\sum_{\vec{k}\neq 0}{\left(e^{\epsilon_{\vec{k}}/T}-1\right)^{-m}}$. The latter formula immediately follows from the formula (\ref{rhoinfinity}) for the unconstrained characteristic function $\Theta^{(\infty)}(u)= \prod_{\vec{k}\neq 0}{\Theta_{n_{\vec{k}}}^{(\infty)}(u)}$ which is a product of the characteristic functions for the independent stochastic variables $n_{\vec{k}}$. Each of the latter functions is easily calculated as an elementary geometric series 
\begin{equation}
\Theta_{n_{\vec{k}}}^{(\infty)}(u) = Tr \{e^{iu\hat{n}_{\vec{k}}} \hat{\rho}_{\vec{k}} \}
= Tr \{\frac{e^{iu\hat{n}_{\vec{k}}} e^{-\varepsilon_{\vec{k}} \hat{n}_{\vec{k}}/T}}
{(1-e^{-\varepsilon_{\vec{k}}/T})^{-1}}\}
=\frac{e^{\epsilon_{\vec{k}}/T}-1}{e^{\epsilon_{\vec{k}}/T}-e^{iu}}.
\end{equation}
\begin{figure}
\centering
\includegraphics[width=13cm]{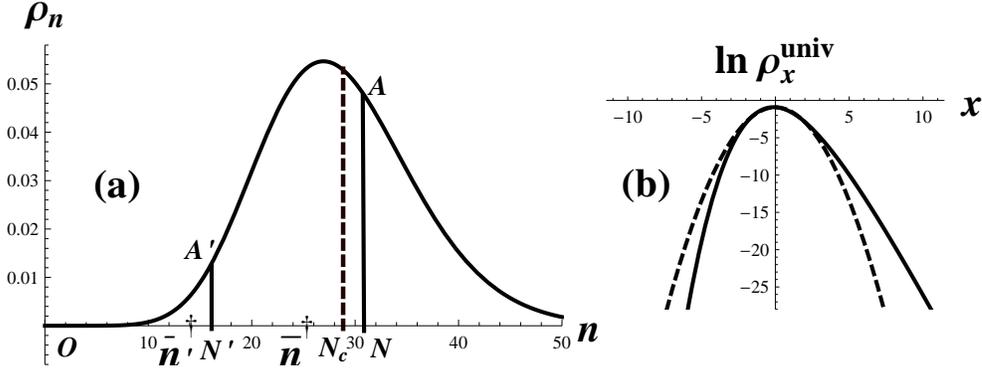}
\caption[example]
{\label{fig1}
(a) Unconstrained probability distribution $\rho_{n}^{(\infty)}$ of the total noncondensate occupation $n$ and its constraint cuts off (solid lines) for a small number of atoms $N' < N_c$ (OA'N' - there is no condensate) and for a large number of atoms $N > N_c$ (OAN - there is condensate) for $N_v = 50$. (b) Logarithm of the universal unconstrained probability distribution (\ref{rhouniv}), $\rho_{x}^{(univ)}$, as a function of the scaled noncondensate occupation $x=(n-N_{c})/\sigma^{(\infty)}$. Logarithm of the Gaussian distribution, $-x^{2}/2- \ln (2\pi) /2$, is depicted by a dashed line.}
\end{figure}
     The solution (\ref{rhocut}) is remarkably simple and nonanalytical, but it is an exact solution. It allows us to solve the problem of critical fluctuations in the BEC in an ideal gas. It is easy to prove that (\ref{rhocut}) is the exact solution to a well-known exact recursion relation \cite{Landsberg,Borrmann1993,Brosens1997,ww1997,Holthaus1997,Balazs1998,recursion1999,Borrmann1999,Kleinert2007,Wang2009}. 
     
\section{Self-similarity of critical BEC fluctuations}
The most efficient way to analyze BEC statistics is to study the central moments and cumulants of the noncondensate occupation as functions of the number of atoms in the trap. These functions are more physically instructive and more directly related to the intrinsic quantum statistics in a finite system than less transparent temperature dependences. BEC phase transition takes place when a number of loaded in the trap atoms reaches a critical number \cite{KKS-PRL,KKS-PRA} given by a discrete sum $N_c = \sum_{\vec{k}\neq 0}{\left(e^{\epsilon_{\vec{k}}/T} - 1\right)^{-1}}$. The latter is equal also to the maximum possible number of noncondensed atoms ${\bar n}^{(\infty)} = N_c$ achieved in the limit of an infinite number of atoms loaded in the trap, $N \to \infty$. In the thermodynamic limit this sum is approximated by a bit larger continuous integral $N_v = \zeta(3/2)[mT/(2\pi \hbar^2)]^{3/2} V = N(T/T_{c})^{3/2}$, where $\zeta$ is the zeta function of Riemann. A ratio of an energy scale for the box trap to the temperature is determined by precisely the same trap-size parameter $N_v$ since the energy of the first excited state is $\epsilon_{1} = \pi T\left[\zeta(3/2)/N_{v}\right]^{2/3}$. Hence, the sum in $N_c$ over the atomic states of the trap as well as all other similar sums, like in (\ref{sigma}) below, actually depend only on a single combination of the trap parameters $N_v$. Thus, the system of the ideal gas atoms in the finite box is completely specified by two parameters, $N_v$ and $N$. It is convenient to study a development of the BEC phase transition with an increase in the number of atoms $N$ assuming that the volume and temperature are fixed, that is the trap-size parameter $N_v$ is fixed. Hence, when we increase the number of atoms from $N < N_c$ to $N > N_c$ the system undergoes the same BEC phenomenon as the one observed when we decrease the temperature around the critical temperature $T_c$ from $T > T_c$ to $T < T_c$. To avoid confusion, we remind that an index $c$ in the critical number of atoms $N_c$ stands for "critical", not for "condensed". The number of condensed atoms is denoted as $n_0$.

    We find the probability distribution of the total noncondensate occupation in the thermodynamic limit, which corresponds to a limit $N_v \to \infty$, using above-formulated constraint-cutoff approach and figure 1a. In that limit the unconstrained probability distribution $\rho_{n}^{(\infty)}$ analytically calculated in \cite{KKS-PRL,KKS-PRA} for different sizes and temperatures of the trap tends to a universal function if it is considered for the scaled stochastic variable $x =(n -N_{c})\epsilon_{1}/(T\sqrt{s_2}) \approx (n -N_{c})/\sigma^{(\infty)}$ centered to have zero mean value. Namely, we find 
\begin{equation}
\frac{T\sqrt{s_2}}{\epsilon_{1}}\rho_{n}^{(\infty)} \to \rho_{x}^{(univ)}=\frac{\sqrt{s_2}}{2\pi}\int_{-\infty}^{\infty}{e^{-iu'x\sqrt{s_2}}\Theta^{(univ)}(u')du'},
\label{rhouniv}
\end{equation}
when $N_v \to \infty$. The universal function is presented here via its Laplace transform, i.e., a characteristic function $\Theta^{(univ)}=\exp [\sum_{m=2}^{\infty}{\frac{s_m}{m}(iu')^{m}}]$. It is immediate to derive the result (\ref{rhouniv}) from the fact that in the thermodynamic limit all higher-order cumulants scale as the powers of the dispersion, $\tilde{\kappa}_{m}^{(\infty)} \to (m-1)! (\sigma^{(\infty)}/s_{2}^{1/2})^m s_m$, where $T/\epsilon_{1} \to \sigma^{(\infty)}/s_{2}^{1/2}$. Indeed, for $m\geq 2$ the main contribution to $\tilde{\kappa}_{m}^{(\infty)}$ comes from the energies much lower than temperature, $\epsilon_{\vec{k}} \ll T$, where $(e^{\epsilon_{\vec{k}}/T}-1)^m \approx (\epsilon_{1}/T)^{m}\vec{k}^{2m}, \vec{k} = 2\pi \vec{q}/L$, and vector $\vec{q}=\{q_{j}, j=x,y,z\}$ has integer components, $q_{j}=0,\pm 1,\pm 2,\ldots$. Here we use the universal numbers $s_m = \sum_{\vec{q}\neq 0}{q^{-2m}}$, given by the generalized Einstein function \cite{KochPhysicaA2001}, and the dispersion of the BEC fluctuations $\sigma^{(\infty)}$ that is independent of the number of atoms $N$ quantity calculated for the unconstrained probability distribution (\ref{rhoinfinity}), 
\begin{equation}
\sigma^{(\infty)}=\sqrt{N_{c}+\sum_{\vec{k}\neq 0}{(e^{\epsilon_{\vec{k}}/T}-1)^{-2}}} \approx \frac{s_{2}^{1/2}}{\pi}\left(\frac{N_{v}}{\zeta(3/2)}\right)^{\frac{2}{3}} = \frac{s_{2}^{1/2}T}{\epsilon_{1}}.
\label{sigma}
\end{equation}
An approximation in (\ref{sigma}) gives a thermodynamic-limit value expressed via the discrete sum $s_2 = \sum_{\vec{q}\neq 0}{q^{-4}} \approx 16.533$. The dispersion of the BEC fluctuations (\ref{sigma}) is anomalously large and scales as $\sigma^{(\infty)} \sim N_{v}^{2/3}$, contrary to a much smaller value $\sim N_{v}^{1/2}$, which one could naively expect from a standard analysis based on the grand-canonical or thermodynamic theory of fluctuations. Please note that in order to pin the critical point to zero, $x_c =0$, we have to measure the scaled variable $x$ relative to the exact mesoscopic critical value $N_c$ given by the discrete sum that cannot be replaced here by its continuous approximation $N_v$. Otherwise, all universal functions for the stochastic and thermodynamic quantities would acquire a trap-size-dependent shift $\eta_{c}=(N_{v}-N_{c})/\sigma^{(\infty)}$ that does not tend to zero, but instead slowly increases like $\sim N_{v}^{\delta}$ with a power index $\delta \sim 0.1$. That shift is not addressed in the usual grand-canonical-ensemble approximation in the thermodynamic limit, but it is important to resolve correctly the universal structure of the critical region as is clearly seen from an example of the heat capacity discussed below in figure 4. From practical calculations' point of view, the universal probability distribution (\ref{rhouniv}) is almost indistinguishable from the exact probability distribution already for the finite-size systems starting from a trap-size parameter $N_v \sim 100$, when a discreteness of the number of atoms in the noncondensate or condensate becomes unimportant. In the latter case, $N_v \sim 100$, one can calculate the probability distribution using function (\ref{rhouniv}) practically for all numbers of atoms in the noncondensate starting from $n > 10$. 

	 We derive also approximate formulas for the universal probability distribution of the total noncondensate occupation in terms of Kummer's confluent hypergeometric function,
\begin{equation}
\rho_{x}^{(univ)}\approx \frac{e_{1}^{g_1}e_{2}^{g_2}X^{g_{1}+g_{2}-1}}{\Gamma (g_{1}+g_{2})e^{e_{2}X}}M(g_{1},g_{1}+g_{2},(e_{2}-e_{1})X),\ X=x+\frac{g_{1}}{e_{1}}+\frac{g_{2}}{e_{2}},
\label{rhounivKummer}
\end{equation}
where $e_{1}\approx 4.303, e_{2}\approx 29.573, g_{1}\approx 8.504, g_{2}\approx 473$, and in terms of the parabolic cylinder function, 
\begin{equation}
\rho_{x}^{(univ)} \approx \frac{c^{g}e^{c^{2}/2-Y^2}}{\sqrt{2\pi (1-\frac{s_{3}^{2}}{s_{2}s_{4}})}}D_{-g}[2(c-Y)],\quad 
Y=\frac{x+\frac{s_{3}\sqrt{s_2}}{s_4}}{2\sqrt{1-\frac{s_{3}^{2}}{s_{2}s_{4}}}},
\label{rhounivPC}
\end{equation}
where $g=s_{3}^{4}/s_{4}^{3}, c=(s_{3}/s_{4})\sqrt{s_{2}-s_{3}^{2}/s_4}$, $s_{3}\approx 8.4019$, $s_{4}\approx 6.9458$. They are obtained by means of an exact analytical solution for a model of an ideal gas in a trap with three energy levels. The upper two levels with energies $e_{1}\epsilon_{1}/s_{2}^{1/2}$ and $e_{2}\epsilon_{1}/s_{2}^{1/2}$ contain $g_1$ and $g_2$ degenerate states, respectively. To find the approximations (\ref{rhounivKummer}) and (\ref{rhounivPC}) we match, using free parameters of the model, the first five or four cumulants of the three-level-trap model with the corresponding cumulants of the universal probability distribution (\ref{rhouniv}), respectively. The exactly solvable models will be discussed elsewhere. The analytical results (\ref{rhounivKummer}) and (\ref{rhounivPC}) are remarkably accurate not only near the critical point, but in the whole central part of the critical region (namely, in the intervals $-4 < x < 10$ and $-3 < x < 6$, respectively). They allow us to calculate analytically the universal functions for all moments of BEC statistics, including the order parameter, and other physical quantities via the exact formulas for the constraint-cutoff distribution (\ref{rhocut}). The analytical solution in terms of Kummer's confluent hypergeometric function (\ref{rhounivKummer}) has wider range of validity, $-4 < x < 10$, covers more than 10 orders in $\rho_{x}^{(univ)}$, and is more accurate, but also more complicated than the solution (\ref{rhounivPC}) in terms of the parabolic cylinder function, which is valid in still very wide interval $-3 < x < 6$, which covers more than 6 orders in $\rho_{x}^{(univ)}$ and includes all of the most interesting for critical phenomena central part of the critical region. Both of these solutions overlap well with the asymptotics (\ref{arhox}) and (\ref{rap}) at the wings of the critical region; thus, both of them yield an analytical solution to the problem of universality of critical fluctuations, Gibbs free energy, heat capacity, and other thermodynamic quantities in the BEC phase transition of an ideal gas. An accuracy of the approximations (\ref{rhounivKummer}) and (\ref{rhounivPC}) is on order of or better than 1 per cent in the whole range of their validity both for the universal function (\ref{rhouniv}) itself and for all thermodynamic quantities discussed below, including heat capacity which is the most sensitive to any small perturbations in the value and derivatives of the probability distribution. 

  	The universal probability distribution $\rho_{x}^{(univ)}$ for the box trap with periodic boundary conditions, according to (\ref{rhouniv}), does not include any parameters and is a pure mathematical special function. A similar universal probability distributions of the total noncondensate occupation can be derived for any other trap, e.g., a box with the Dirichlet boundary conditions, following exactly the same scheme starting from the unconstrained probability distribution $\rho_{n}^{(\infty)}$ known for an arbitrary trap \cite{KKS-PRL,KKS-PRA}. That remarkable universality is valid in the whole critical region, $\left|n-N_{c}\right|/\sigma^{(\infty)} \ll N_{v}^{1/3}$, which tends to an infinite interval of values $x\in (-\infty , \infty)$ in the thermodynamic limit $N_v \to \infty$. It includes very large values $|x| \gg 1$ and is much wider than a relatively narrow vicinity of the maximum of the distribution, $\left|n - N_{c}\right| \leq \sigma^{(\infty)}$, where a Gaussian approximation always works well. Thus, the actual mesoscopic probability distribution becomes very close to the universal, thermodynamic-limit probability distribution and clearly reveals an intrinsic critical structure of the BEC statistics already starting from quite moderate values of the trap-size parameter $N_v \sim 10^2$. The universal probability distribution has very fat and long right tail (\ref{rap}) of large occupation values, $n - N_{c} \gg \sigma^{(\infty)}$, whereas the left tail (\ref{arhox}) of small occupation values, $N_{c} - n \gg \sigma^{(\infty)}$, is strongly suppressed compared to the Gaussian distribution (see figure 1b). The universal probability distribution does not collapse to a pure Gaussian distribution but remains nontrivial in the thermodynamic limit. We find that a Taylor series for the logarithm of the universal probability distribution, i.e., the negative Landau function 
$\ln \rho_{x}^{(univ)} = a_0 + a_{2}(x-\Delta x)^{2}/2!+ a_{3}(x - \Delta x)^{3}/3! + a_{4}(x - \Delta x)^{4}/4! + \ldots$, contains very essential third ($a_3 \approx 0.26$) and fourth ($a_4 \approx -3$) order terms which provide the same or larger contributions at the tails compared to that of the quadratic part ($a_2 \approx -1.04$). There is a finite shift $\Delta x \approx -0.12$ of the maximum of the probability distribution to the left of the mean value $\bar x =0$, i.e. ${\bar n} = N_c$, due to the above discussed asymmetric tails. The normalization coefficient is $a_0 \approx -\ln \sqrt{2\pi} + 0.013$. Note that a pure Gaussian distribution has only two nonzero coefficients, $a_{0}^{(Gauss)} = -\ln \sqrt{2\pi} \approx -0.919$ and $a_{2}^{(Gauss)} = -1$.
 
 \section{Asymptotics of BEC statistics} 
Asymptotics of the universal probability distribution (\ref{rhouniv}) in the noncondensed and condensed phases is strongly asymmetric:
\begin{equation}
\rho_{x}^{(univ)} \approx \frac{s_{2}^{3/4}(x_0 -x)^{1/2}}{2\pi^{5/2}}e^{f_{0}+\frac{s_{2}^{3/2}}{12\pi^4}(x-x_{0})^{3}},\quad -x\gg 1,
\label{arhox}
\end{equation}
\begin{equation}
\rho_{x}^{(univ)} \approx \sqrt{s_2}\alpha_{5}(x)e^{-\sqrt{s_2}x-g_{1}+s_{0}'},\quad x\gg 1,
\label{rap}
\end{equation}
\begin{equation}
\alpha_{5}(x) = \frac{x_{1}^5}{5!}+\frac{s_{2}'x_{1}^3}{12}-\frac{s_{3}'x_{1}^2}{6}+\left(\frac{s_{2}'^2}{2}+s_{4}'\right)\frac{x_1}{4}-\frac{s_{2}'s_{3}'}{6}-\frac{s_{5}'}{5},
\label{alpha5}
\end{equation}
where $x_{0}\approx 2.2$, $f_{0}\approx 3.3$, $g_{1}=6$ is degeneracy of the first excited energy level in the box trap, $x_1 =\sqrt{s_2}x +g_1 -s_{0}''$, $s_{0}' =\sum_{m=2}^{\infty}{\frac{s_{m}-6}{m}} \approx 6.45$,
$s_{0}'' =\sum_{\left\{\vec{q}: |\vec{q}|>1\right\}}{q^{-2}/(q^{2}-1)} \approx 14.7$, $s_{j}' =\sum_{\left\{\vec{q}: |\vec{q}|>1\right\}}{(q^{2}-1)^{-j}}$ for $j = 2, 3, \ldots$. Here the sum $\sum_{\left\{\vec{q}: |\vec{q}|>1\right\}}$ runs over cubic lattice of all wave vectors $\vec{q} = \{q_{x}, q_{y}, q_{z}\}$ with integer components $q_{x,y,z} = 0, \pm 1, \pm 2, \ldots$ of all atomic states, excluding ground and first excited energy levels ($|\vec{q}|\neq 0, 1$). The most striking result is an incredibly fast cubic exponential decay at the left tail (noncondensed phase, see (\ref{arhox})) and very slow linear exponential decay at the right tail (condensed phase, see (\ref{rap})). Both decays are quite different from the standard in statistical physics Gaussian, quadratic exponential decay. 

\section{Self-similarity and universal structure of the BEC order parameter}
To resolve the universal structure and self-similarity  of the BEC order parameter, we divide both the function and the argument by the dispersion (\ref{sigma}) of the BEC fluctuations $\sigma^{(\infty)}$ and calculate a scaled condensate occupation ${\bar n}_{0}' = {\bar n}_0 /\sigma^{(\infty)}$ as a function of a scaled deviation from the critical point $\eta = (N- N_c )/\sigma^{(\infty)}$. We find that with an increase in the trap-size parameter $N_v$ the function ${\bar n}_{0}'(\eta )$ quickly converges to a universal function 
\begin{equation}
F_{0}(\eta ) = \eta - \int_{-\infty}^{\eta}{\frac{x\rho_{x}^{(univ)}dx}{P_{\eta}^{(\infty)}}}, P_{\eta}^{(\infty)}=\int_{-\infty}^{\eta}{\rho_{x}^{(univ)}dx}, \eta = \frac{N- N_c}{\sigma^{(\infty)}}, 
\label{n0eta}
\end{equation}
\begin{figure}
\centering
\includegraphics[width=10cm]{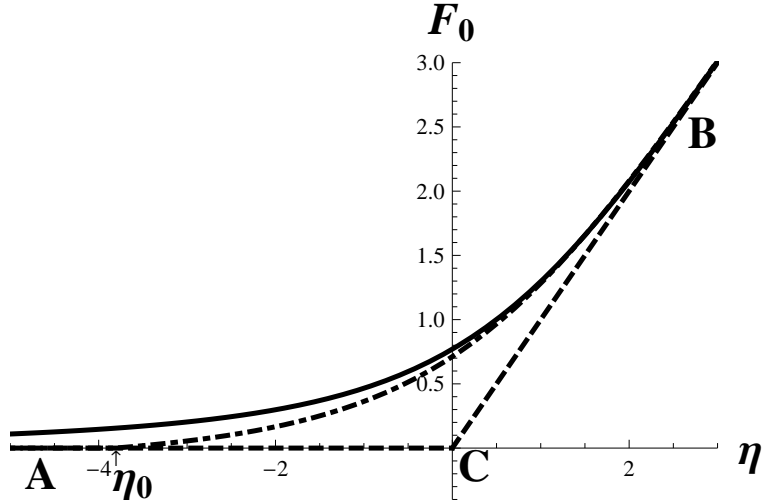}
\caption[example]
{\label{fig2}
The structure of the scaled BEC order parameter ${\bar n}_{0}' = {\bar n}_0 /\sigma^{(\infty)}$ as a function of $\eta = (N- N_c )/\sigma^{(\infty)}$ in the critical region: the solid line is the universal function $F_{0}(\eta)$ and the dotted-dashed line is the function ${\bar n}_{0}'(\eta )$ for the mesoscopic system with the trap-size parameter $N_v = 50$. The angle ACB represents the prediction of the standard Landau mean-field theory.}
\end{figure}
which describes the universal structure of the BEC order parameter in the critical region in the thermodynamic limit, as shown in figure 2. Explicit formula (\ref{n0eta}) for the universal function $F_{0}(\eta )$ immediately follows from the  formula for the exact, cutoff probability distribution (\ref{rhocut}) and from the universal probability distribution (\ref{rhouniv}).

     The exact analytical result (\ref{n0eta}) for the universal structure of the BEC order parameter in the critical region can be easily written in terms of the polynomial, exponential, confluent hypergeometric, and parabolic cylinder functions if we use the explicit formulas for $\rho_{x}^{(univ)}$ in (\ref{rhounivKummer}) and (\ref{rhounivPC}) for the central part of the critical region and in (\ref{arhox}) and (\ref{rap}) for the left (condensed) and right (noncondensed) wings of the critical region to calculate an explicit integral in (\ref{n0eta}). We skip these straightforward expressions in order to not overload the paper with formulas. The universal function of the BEC order parameter $F_{0}(\eta)$ is truly universal since it does not contain any physical parameters of the system and involves only pure numbers $\pi$ and $s_m$. The result (\ref{n0eta}) is very different from the prediction of the Landau mean-field theory shown by the broken line ACB in figure 2. We can immediately conclude that, starting from small mesoscopic systems with $N_v \sim 10^2$, the difference between the universal order-parameter and the mesoscopic order-parameter functions is relatively small, $\left|F_{0}(\eta ) \sigma^{(\infty)} - {\bar n}_{0}(\eta )\right| \ll {\bar n}_{0} (\eta )$. This statement is true everywhere except at the very beginning of the curve ${\bar n}_{0}(\eta )$, where the system is not mesoscopic anymore and there are only a few atoms in the trap $N \leq 10$. Obviously, the number of atoms in the condensate should become exactly zero, ${\bar n}_0 = 0$, at the end point $N=0$, i.e. at $\eta_0 (N_{v}) = -N_{c}/\sigma^{(\infty)} \sim -N_{v}^{1/3}$, as is seen in figure 2 at $\eta_0 \approx -3.817$ for $N_v = 50$. With further increase in the trap-size parameter $N_v$, the end point quickly tends to negative infinity, $\eta_0 \to -\infty$, and the scaled order parameter curve ${\bar n}_{0}'(\eta )$ merges the universal function $F_{0}(\eta)$. At the critical point, where the number of atoms in the trap is critical, $N = N_c$, the order parameter just reaches a level of fluctuations, ${\bar n}_0 \approx 0.77 \sigma^{(\infty)}$. The standard Landau mean-field theory does not resolve the smooth, regular universal structure in figure 2.
    
       We stress that the usual grand-canonical-ensemble approximation fails \cite{Koch06,Pathria} in the whole critical region, $\left|N_c - N\right| \leq 2\sigma^{(\infty)}$, and in the region of a well-developed BEC, $N - N_c \gg 2\sigma^{(\infty)}$. It is valid only in the limit of the small number of atoms, $N_c - N \gg \sigma^{(\infty)}$, which corresponds to a high-temperature regime of a classical gas without condensate. The only excuse for the grand-canonical-ensemble approximation utilizing a pure exponential distribution is its simplicity.
       
\section{Self-similarity and universal structure of higher-order cumulants and moments}
All higher-order moments and cumulants of the BEC fluctuations also demonstrate self-similarity and universal structure. That fact immediately follows from the universality of the noncondensate occupation probability distribution (\ref{rhouniv}). The analysis is similar to the one developed above for the order parameter and is based on the calculation of the scaled central moments $\mu_{m}' = \mu_{m}/(\sigma^{(\infty)})^m$ and scaled cumulants $\kappa_{m}' = \kappa_{m}/(\sigma^{(\infty)})^m$ as functions of the scaled deviation from the critical point, $\eta = (N- N_c )/\sigma^{(\infty)}$. We find that with an increase in the trap-size parameter $N_v$ the function $\mu_{m}'(\eta )$ quickly converges to the universal function
\begin{equation}
M_{m}(\eta ) = \int_{-\infty}^{\eta}{(x-\bar{x})^{m} \rho_{x}^{(univ)}dx}/P_{\eta}^{(\infty)},
\label{Meta} 
\end{equation}   
and the function $\kappa_{m}'(\eta )$ - to the corresponding function $C_{m}(\eta )$. 
\begin{figure}
\centering
\includegraphics[width=10cm]{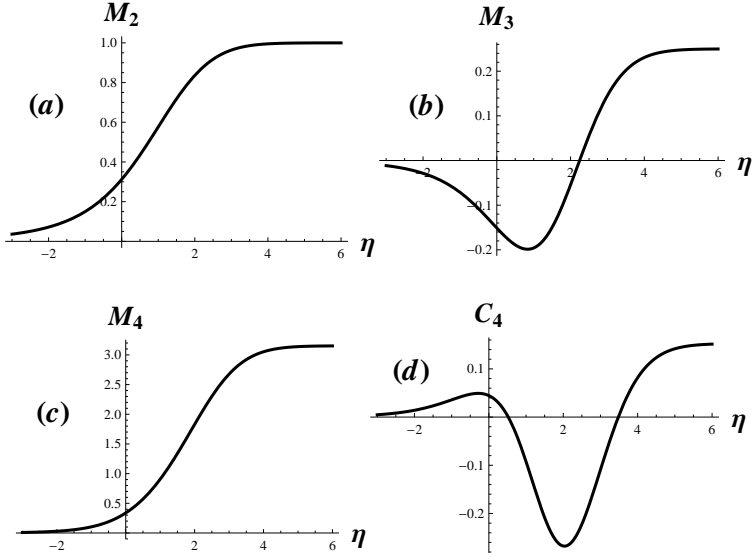}
\caption[example]
{\label{fig3}
Universal functions $M_{m}(\eta)$ and $C_{m}(\eta)$ of the scaled central moments and cumulants (a) $\mu_{2}' \equiv \kappa_{2}' = \mu_{2}/(\sigma^{(\infty)})^2$, (b) $\mu_{3}' \equiv \kappa_{3}' = \mu_{3}/(\sigma^{(\infty)})^3$, (c) $\mu_{4}' = \mu_{4}/(\sigma^{(\infty)})^4$, and (d) $\kappa_{4}' \equiv \mu_{4}' - 3(\mu_{2}')^2$ of the total noncondensate occupation in the critical region as functions of $\eta = (N- N_c )/\sigma^{(\infty)}$.}
\end{figure}
One can easily write down explicit formulas for the universal functions $F_{0}(\eta ), M_{m}(\eta )$, and $C_{m}(\eta )$ of the order parameter, central moments, and cumulants as the simple integrals in (\ref{n0eta}) and (\ref{Meta}) via the universal unconstrained probability distribution $\rho_{x}^{(univ)}$ given by the explicit analytical formulas (\ref{rhounivKummer}) and (\ref{rhounivPC}) (the central part of the critical region) and (\ref{arhox}) and (\ref{rap}) (asymptotics of the left and right wings of the critical region). 

     These functions describe the universal structure of the BEC critical fluctuations in the thermodynamic limit, $N_v \to \infty$, as shown in figure 3 for the second, third, and fourth moments and cumulants of the noncondensate occupation. The above explicit formulas for these universal functions immediately follow from the exact formulas (\ref{rhocut}) and (\ref{rhouniv}), similar to the derivation of (\ref{n0eta}). The functions $M_{m}(\eta )$ and $C_{m}(\eta )$ do not involve any physical parameters and are truly universal. The universal behaviour is clearly observed starting from very small mesoscopic systems, $N_v \sim 100$. An essential deviation from the universal curves takes place only near the end point, $\eta_0$, where the number of atoms in the trap is zero and, hence, all fluctuations $\mu_{m}'(\eta )$ and $\kappa_{m}'(\eta )$ should be exactly zero, as is seen in figure 2 at $\eta_0 \approx -3.817$ for $N_v = 50$. Near the end point $\eta_0$, where the system loses its mesoscopic status, it can be studied quantum mechanically as a microscopic system of a few atoms $N = 1, 2, 3, \ldots$. 
     
     Behaviour of the moments and cumulants, depicted in figure 3, can be qualitatively predicted on the basis of the constraint-cutoff mechanism using figure 1a. The variance $\mu_2 \equiv \kappa_2$ has to grow monotonically with increasing $N$ and to have a maximum derivative at $N \approx N_c$ because the width of the cutoff probability distribution OAN increases when the cutoff boundary AN moves to the right and the maximum width's derivative is achieved at the center of the critical region. That behaviour, indeed, is found in the universal function $M_{2}(\eta)$ depicted in figure 3a. The third central moment, or the third cumulant, $\mu_3 \equiv \kappa_3 =\left\langle (n - {\bar n})^{3}\right\rangle$, is the main characteristic of an asymmetry of the probability distribution relative to the mean value $\bar n$. For small enough numbers of atoms in the trap $N$, when the probability distribution has a strongly asymmetric, "curved-triangle" shape OA'N' in figure 1a, the value of the asymmetry $\mu_3$ is negative due to a large contribution from the left tail and increases in magnitude with increasing $N$ until some maximum-in-magnitude negative value is reached. When the number of atoms $N$ enters the central part of the critical region, the absolute value of the asymmetry $\left|\mu_3\right|$ decreases and after passing through the critical point $N = N_c$ approaches zero, since the shape of the cutoff probability distribution OAN in figure 1a becomes more and more symmetric. Finally, the asymmetry coefficient $\mu_{3}' = \mu_{3}/(\sigma^{(\infty)})^3$ changes the sign and tends to a finite positive value $\mu_{3}'^{(\infty)} = 0.25$ for $N_v \to \infty$, which is a characteristic feature of the unconstrained distribution $\rho_{n}^{(\infty)}$ due to a large positive contribution of the fat and wide right tail, discussed in figure 1b. The universal function $M_{3}(\eta )=C_{3}(\eta )$ in figure 3b, indeed, follows that behaviour of the asymmetry $\mu_3 = \kappa_3$. 
     
     Depicted in figures 3c,d behaviour of the fourth moment $\mu_4$ and the fourth cumulant, i.e., the excess $\kappa_4$, also can be easily explained. The fourth cumulant, in general, characterizes a positive excess (if $\kappa_4 > 0$) or a deficit (if $\kappa_4 < 0$) of the flatness of the "plateau" of the probability distribution relative to the flatness of the plateau of the Gaussian distribution. Again, one has to take into account that the unconstrained probability distribution $\rho_{n}^{(\infty)}$, according to figure 1b, is more flat than the Gaussian distribution, that is, it has a positive excess coefficient $\kappa_{4}'^{(\infty)} \approx 0.15$ for $N_v \to \infty$. 
     
    In the low-temperature regime of the fully developed condensate, the cutoff part of the probability distribution in figure 1a contains only an unimportant end piece of the right tail. Thus, the mean value as well as all moments and cumulants tend to the constants, which are precisely their unconstrained values analytically calculated in \cite{KKS-PRL,KKS-PRA}. In particular, the limiting values of the scaled cumulants are equal to $\kappa_{m}'^{(\infty)} =\tilde{\kappa}_{m}^{(\infty)}/(\sigma^{(\infty)})^{m} =(m-1)!s_{m}/s_{2}^{m/2}$ so that the asymmetry and excess coefficients tend to $\mu_{3}'^{(\infty)} =2s_{3}/s_{2}^{3/2} \approx 0.25$ and $\kappa_{4}'^{(\infty)} = 6s_{4}/s_{2}^{2} \approx 0.15$, respectively. We find that this is indeed true, as is clearly seen in figures 2 and 3. 
     
\section{Self-similarity and universal structure of the thermodynamic quantities} 
Thermodynamics in the whole critical region around the $\lambda$ point can be immediately found and resolved on the basis of the above analytical theory of BEC statistics. Let us start with the Gibbs free energy \cite{LLV} $F = -T\ln Z$, which is determined by the partition function $Z =Tr\left\{e^{-{\hat H}/T}\theta (N-{\hat n})\right\}$. The Gibbs free energy is the basic, generating function for the thermodynamics in the canonical ensemble of particles since its derivatives determine the main thermodynamic quantities, including the average energy $\bar{E}=F+TS$, entropy $S=-\left(\partial F/\partial T\right)_V$, and heat capacity $C_V =\left(\partial\bar{E}/\partial T\right)_V$. We find that with increasing trap-size parameter $N_v$ the Gibbs free energy in the critical region is determined by the universal function of $\eta$,
\begin{equation}
\frac{F - F^{(\infty)}}{T}=-\ln P_{\eta}^{(univ)},\quad P_{\eta}^{(univ)}=\int_{-\infty}^{\eta}{\rho_{x}^{(univ)}dx},
\label{Gibbsuniv}
\end{equation}
where the cumulative distribution function $P_{\eta}^{(univ)}$ can be easily calculated from the above-presented universal probability distribution $\rho_{x}^{(univ)}$. The term $F^{(\infty)} =T\sum_{\vec{k}\neq 0}{\ln \left(1-e^{-\epsilon_{\vec{k}}/T}\right)}$ is a well-known in statistical physics \cite{LLV} contribution from the unconstrained noncondensate excitations which is independent of the number of atoms and tends to $-TN_{v}\zeta(5/2)/\zeta(3/2)$ in the thermodynamic limit. 

    Similarly, the average energy is determined by the universal function 
\begin{equation}
\frac{\bar{E}-\bar{E}^{(\infty)}}{\sqrt{\sigma^{(\infty)}}T}= -\frac{3\pi^{3/2}\zeta (3/2)\rho_{\eta}^{(univ)}}{2s_{2}^{3/4}P_{\eta}^{(univ)}},\ \bar{E}^{(\infty)}=\sum_{\vec{k}\neq 0}{\frac{\epsilon_{\vec{k}}}{e^{\epsilon_{\vec{k}}/T}-1}},
\label{Euniv}
\end{equation}
where $\bar{E}^{(\infty)}$ is a well-known contribution from the unconstrained noncondensate excitations which tends to $-(3/2)F^{(\infty)}$ in the thermodynamic limit. 

     The most often addressed thermodynamic quantity in the studies of BEC phase transition is the specific heat $c=C_{V}/N$, that is the heat capacity per particle. It is directly measured in experiments and has a subtle structure near the critical point resembling the Greek letter $\lambda$ (hence the name $\lambda$ point for the critical point). The $\lambda$-point structure of the specific heat of an ideal gas can be described by a universal function that surprisingly has not yet been analytically found despite many studies since the original works by Bose and Einstein in 1924. An explicit analytical solution to that long-standing problem is given below.
     
     The standard grand-canonical-ensemble approximation for the value of the specific heat at the critical point $c_{0} =N_{c}^{-1}\sum_{\vec{k}\neq 0}{(\epsilon_{\vec{k}}/T_{c})^{2}e^{\epsilon_{\vec{k}}/T_{c}}(e^{\epsilon_{\vec{k}}/T_{c}}-1)^{-2}}$ tends, in the thermodynamic limit, to a well-known value $(15/4)\zeta(5/2)/\zeta(3/2) \approx 1.92567$ which is larger than the classical gas specific heat $3/2$. To resolve the $\lambda$-point structure, we have to calculate correctly the terms of the next order $\sim 1/\sqrt{\sigma^{(\infty)}} \sim N_{v}^{-1/3} \ll 1$. The result for the specific heat at the critical point $\eta =0$ is $c_{c} =c_{0} + c_{1}/\sqrt{\sigma^{(\infty)}}$, where $c_{1} =(9/4)\pi^{\frac{3}{2}}\zeta(\frac{3}{2})s_{2}^{-\frac{3}{4}}[\partial^{2}\ln P_{\eta}^{(univ)}/\partial\eta^{2}]_{\eta=0} \approx -2.79$. For large trap-size parameter $N_v \gg 1$, the specific heat in the critical region is determined by the universal function $F_{C}^{(univ)}(\eta) =\sqrt{\sigma^{(\infty)}}(c-c_{c})$. We find 
\begin{equation}
F_{C}^{(univ)}(\eta)=\frac{9\pi^{\frac{3}{2}}\zeta(\frac{3}{2})}{4s_{2}^{3/4}}\frac{\partial^{2}\ln P_{\eta}^{(univ)}}{\partial \eta^2}-c_{1}-\frac{15\zeta(\frac{5}{2})s_{2}^{3/4}\eta}{4\pi^{\frac{3}{2}}(\zeta(\frac{3}{2}))^2}.
\label{FCVuniv}
\end{equation}
It is depicted in figure 4a and has a familiar $\lambda$ shape. Its asymptotics, 
\begin{equation}
F_{C}^{(univ)}\approx \frac{s_{2}^{3/4}9\zeta(\frac{3}{2})\theta(-\eta)\eta}{8\pi^{\frac{5}{2}}}-\frac{s_{2}^{3/4}15\zeta(\frac{5}{2})\eta}{4\pi^{\frac{3}{2}}(\zeta(\frac{3}{2}))^2}-c_{1},\quad |\eta|\gg 1,
\label{FCVunivap}
\end{equation}
\begin{figure}
\centering
\includegraphics[width=12.5cm]{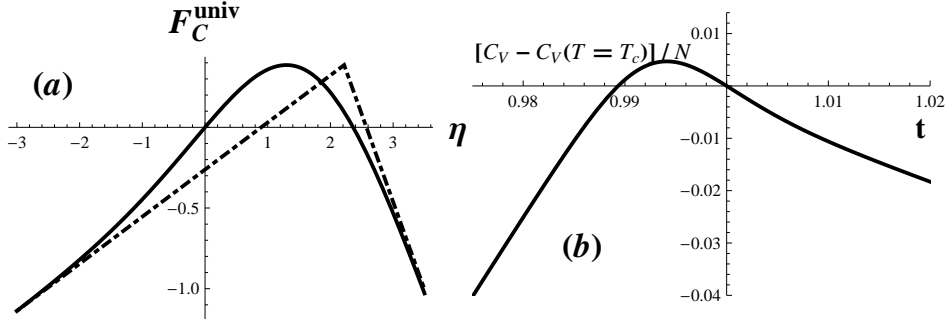}
\caption[example]
{\label{fig4}
(a) Universal function $F_{C}^{(univ)}(\eta)$, (\ref{FCVuniv}), of the scaled specific heat as a function of $\eta = (N - N_{c})/\sigma^{(\infty)}$ in the critical region. The slopes of its linear asymptotics (\ref{FCVunivap}) on both wings of the critical region match the ones predicted by the standard grand-canonical-ensemble approximation. (b) Deviation of the specific heat from its critical value, $[C_{V} - C_{V}(T=T_{c})]/N$, as a function of reduced temperature $t=T/T_c$. The graph is plotted by means of the scaled universal function $F_{C}^{(univ)}(\eta (t))/\sqrt{\sigma^{(\infty)}(t)}$ in (\ref{FCVuniv}) and self-similar substitution (\ref{etat}) for the mesoscopic system with $N=10^6$ atoms in the trap.}
\end{figure}
follows from (\ref{arhox}) and (\ref{rap}). Both slopes of the asymptotics, shown in figure 4a, are exactly the same as the ones given by the standard grand-canonical-ensemble approach. The result (\ref{FCVunivap}) predicts exact positions of these linear asymptotic lines relative to the critical point. These positions were not given right by the grand-canonical-ensemble approach \cite{LLV,Pathria}, which does not resolve the fine universal structure of the $\lambda$ point reducing it, instead, to only a discontinuity with a jump $(9/(8\pi))(\zeta(3/2))^{2}\sigma^{(\infty)}/N_{v}$ in its first derivative with respect to $\eta$ at the critical point and gives correctly only the slopes of the asymptotics and the maximum value in the main order, $c_0$. It does not predict correctly the shifts of the asymptotics and the shift of the maximum value as well as the whole fine structure in the vicinity of the $\lambda$ point which are all quantities of the next order of magnitude $\sim N_{v}^{-1/3}\ll 1$.

\section{Self-similar scaling of statistical and thermodynamic quantities}
The existence of the universal functions for all statistical and thermodynamic quantities in the critical region yields a remarkable conclusion that all these quantities depend on the only universal variable $\eta =(N-N_{c})/\sigma^{(\infty)}$. It means definite self-similarity of all curves for a given statistical or thermodynamic quantity as a function of any physical variable at fixed values of other parameters of the system. For example, we can immediately find and plot the dependence of the specific heat on temperature $t=T/T_c$ in the critical region at any particular values of the number of atoms in the trap $N$ and volume $V$ as the function $F_{C}^{(univ)}(\eta(t))/\sqrt{\sigma^{(\infty)}(t)}$ where the self-similar variable $\eta (t)=(N-N_{c}(t))/\sigma^{(\infty)}(t)$ depends on temperature via the dependences of the critical number $N_c(t)$ and the dispersion $\sigma^{(\infty)}(t)$, (\ref{sigma}), on the reduced temperature $t=T/T_c$. It is especially simple if one does not need to take into account relatively small finite-size effects of deviation of the values of the exact discrete sums for the critical number $N_c$ and for the dispersion $\sigma^{(\infty)}$ in (\ref{sigma}) from their continuous approximations. In this case we have the simple self-similarity 
\begin{equation}
N_{v}(t) = Nt^{\frac{3}{2}},\ \sigma^{(\infty)}(t) =\frac{s_{2}^{\frac{1}{2}}N^{\frac{2}{3}}t}{\pi \zeta(\frac{3}{2})^{\frac{2}{3}}},\
\eta(t) =\frac{\pi \zeta(\frac{3}{2})^{\frac{2}{3}}N^{\frac{1}{3}}(1-t^{\frac{3}{2}})}{s_{2}^{\frac{1}{2}}t},
\label{etat}
\end{equation}
where we assume that $N_v \gg 1$ and $|t-1| \ll 1$. Hence, the temperature dependence of the specific heat near the $\lambda$ point is fully specified by the function $F_{C}^{(univ)}(\eta(t))/\sqrt{\sigma^{(\infty)}(t)}$ for any particular values of the number of atoms in the trap $N$ as shown in figure 4b for $N = 10^{6}$. That shape of the specific heat near the $\lambda$ point is the same as the one numerically calculated in \cite{Kleinert2007} from the exact recursion relation. It is worth noting that for such still not very large number of atoms there is noticeable shift of the critical point on the order $-\eta_{c}=(N_{v}-N_{c})/\sigma^{(\infty)}$. That finite-size effect is well known from numerical calculations (see nice graphs in \cite{Kleinert2007}) and can be easily taken into account via the accurate value of $N_c$ in the self-similar variable $\eta (t)$. For example, in the case of $N=N_v = 10^4$ we have $N_c \approx 8663$ and $\sigma^{(\infty)} \approx 300$, so that $-\eta_c \sim 4$ and the critical temperature shift is $\Delta t_c \sim 0.1$. That same self-similarity can be used to find and plot temperature dependences of all other statistical and thermodynamic quantities in the critical region at particular values of the number of atoms $N$ and volume $V$ from their universal functions, e.g., temperature dependence of the BEC order parameter on the basis of its universal function (\ref{n0eta}). There has been a large amount of papers devoted to the numerical studies of various dependences of the statistical and thermodynamic quantities at numerous possible combinations of the parameters in the finite-size systems (see, e.g., \cite{Koch06,Ziff,Kleinert2007,KKS-PRA,CNBII,Ketterle1996,BagnatoPRA1997}). Nevertheless, the above self-similarity, even in the continuous approximation (\ref{etat}), has not been discovered although has been present in those simulations.

\section{Comparison with renormalization-group theory and critical exponents}
Presented above analytical microscopic theory of the second-order phase transitions yields the full quantum-statistical description of the critical fluctuations phenomena and allows us to find both universal quantities (critical exponents) and nonuniversal quantities (in particular, scaling functions and metric amplitudes) which were introduced in the renormalization-group theory for a close vicinity of the critical point (see [4, 9-14] and references therein). It is known \cite{Fisher1972,Fisher1973,Fisher1974,Fisher1986,Pathria} that BEC phase transition in an ideal gas trapped in the box belongs to the universality class of the Gaussian complex-field model (spherical model) and, hence, the correlation-length exponent $\nu$ is equal to the condensate-fraction exponent $\textit{v}$, $\nu = \textit{v} = 1$, and the specific-heat exponent is $\alpha = -1$. The BEC phase transition in a weakly interacting Bose gas belongs to the three-dimensional XY, or O(2), universality class \cite{Fisher1986,Kleinert1989,Pollock1992,Schultka1995,Wang2009}, which has different critical exponents ($\nu \approx 0.6717$ and $\alpha \approx -0.015$, see \cite{Svistunov2006,Campostrini2009,Gasparini} for the most recent numerical and experimental data) and will be discussed elsewhere. The specific-heat and correlation-length exponents are related via the hyperscaling relation $\alpha = 2 - d \nu$ and the condensate-fraction (superfluid-stiffness) and correlation-length exponents are related via the Josephson scaling relation $\textit{v} = (d-2)\nu$, so that $\textit{v} = (d-2)(2-\alpha)/d$, where $d$ is the dimensionality. For the box trap $d$ is equal to 3. 
     
     The modern theory of the second-order phase transitions, that is the renormalization-group theory, analyzes the finite-size scaling in the close vicinity of the critical point $T = T_c$ on the basis of a power-law ansatz for a critical ("singular") part of a physical quantity as a function of the reduced temperature $\Delta t = (T-T_{c})/T_c$ and the size of the system $L = V^{1/3}$. In the present case of the BEC phase transition, for any physical quantity it is convenient to introduce its properly normalized value, say $y(\Delta t,L)$, which is finite in the thermodynamic limit, $y(0,\infty) = y^{(c)}$ at $L \to \infty$, at the critical point $T = T_c$ , or $N = N_{c}$. In this case, the renormalization-group ansatz  \cite{Schultka1995,Privman1984,Pollock1992,Gasparini,Campostrini2009} reads 
\begin{equation}
y(\Delta t, L, N) = y^{(c)} + \left|\Delta t\right|^{\zeta_y} g_{y}\left[L/\xi (\Delta t)\right],
\label{RGansatz}
\end{equation}
where $\zeta_y$ and $g_y$ are the critical exponent of the physical quantity $y$ and its universal scaling function, respectively, and $\xi(\Delta t) \equiv \xi(\Delta t, L=\infty) =\xi_0 |\Delta t|^{-\nu}$ is the correlation length $\xi(\Delta t, L) = |\Delta t|^{-\nu} f_{\xi}(L/\xi(\Delta t))$ at $L = \infty$, i.e., for the infinite-size system.    
     
     Let us take the condensate fraction ${\bar n}_{0}/N$ or specific heat $c = C_{V}/N$ as the physical quantity $y$. As we find in the previous sections, such quantities are actually described by the universal functions of the self-similar variable $\eta = (N-N_{c})/\sigma^{(\infty)}$ rather than the renormalization-group scaling variable $L/\xi(\Delta t)$. However, these two variables are in fact proportional to each other in the close vicinity of the critical point for large enough systems (to the first order), $\eta \approx -(3/2)(N_{v}/\sigma^{(\infty)})\Delta t \propto \Delta tL$. In a general case, we have $\sigma^{(\infty)} \propto L^{\Delta_{\sigma}}$, where the scaling dimension $\Delta_{\sigma}$ of the dispersion $\sigma^{(\infty)}$ depends on the trap and is equal to 2 for box and to 3/2 for harmonic trap \cite{KKS-PRL,KKS-PRA,Koch06}. So, the exact universal structure of any physical quantity $y$ in the critical region, which is described by the appropriate (found above) universal function of the true universal variable $\eta$ and can be written as 
\begin{equation}
y(T, L, N) = y^{(c)} + \left(\sigma^{(\infty)} \right)^{-\zeta_y /(\nu \Delta_{\sigma})} f_{y}(\eta) , 
\label{eta-ansatz}
\end{equation}
is reduced to the renormalization-group ansatz (\ref{RGansatz}) very close to the critical point. As a consequence, the correlation-length exponent can be found just from the knowledge of the true self-similar variable $\eta$, without any specific information on the universal functions. 

     The analytical microscopic theory in (\ref{n0eta}) yields the universal scaling of the condensate fraction ${\bar n}_{0}(T, L, N)/N = \left(\sigma^{(\infty)} \right)^{-\textit{v}/(\nu \Delta_{\sigma})} f_{n_{0}}(\eta)$ with the explicit formula for the condensate universal function as a regular function of $\eta$ and the critical exponent $\textit{v} = 1$ derived from (\ref{sigma}). It is exactly the same value of the critical exponent that can be directly obtained via the Josephson scaling relation $\textit{v} = (d-2)\nu$ from the correlation-length critical exponent $\nu = 1$ derived from $\eta\propto \Delta tL$. For the specific heat, the analytical microscopic theory yields the universal scaling $c(T, L, N) =  c_c + \left(\sigma^{(\infty)} \right)^{\alpha /(\nu \Delta_{\sigma})} f_{Cv}(\eta)$ with the explicit formula for the specific-heat universal function (\ref{FCVuniv}) and the critical exponent $\alpha = -1$. 

	   Phenomenological renormalization-group theory does not give any explicit formulas for the universal functions of the above discussed and other physical quantities, but usually evaluates them on the basis of numerical, first of all Monte Carlo, simulations and some numerical fits for the few first terms in their Taylor series including corrections from some irrelevant scaling field (see, e.g., \cite{Fisher1986,Pollock1992,Schultka1995,Ceperley1997,Holzmann1999,Campostrini2009,Wang2009} and references therein). That procedure has certain problems in providing the form of the universal functions relatively far from the critical point, i.e., in the whole critical region, and with high enough accuracy. As we discussed above, the universal functions are highly nontrivial functions even in the case of an ideal gas and the true universal, self-similar variable $\eta$ is different from the usually assumed one, $\Delta t L^{1/\nu}$. Besides, such direct simulations of the finite-size versions of the universal functions for relatively small systems are greatly subject to finite-size effects which are difficult to separate from the universal part of the functions without knowing the universal constrain-cutoff mechanism that basically controls the critical phenomena in the second-order phase transitions. All these reasons prevented the renormalization-group approach from finding the full fine structure of the universal functions in the whole critical region. 
     
     In the present paper we outlined analytical microscopic theory of the second-order phase transitions. It yields the explicit formulas for the true universal, self-similar variable and the universal functions such as the ones given in (\ref{n0eta})-(\ref{FCVuniv}). They describe a nontrivial structure of the whole critical region in complete detail, not just the linear terms in the very vicinity of the critical point which were usually discussed and numerically fitted in the renormalization-group analysis. The analytical microscopic and renormalization-group theories coincide only in the first order near the critical point where the true universal variable $\eta$ is reduced to the finite-size scaling variable $L/\xi(\Delta t) \propto \Delta tL$. It is worth noting that a deviation of any physical quantity, scaled with a positive critical exponent, $\zeta_y > 0$, from its thermodynamic-limit critical value $y^{(c)}$ in general is not zero as one could naively expect from the renormalization-group ansatz $|\Delta t|^{\zeta_y} g_{y}(L/\xi(\Delta t))$ at $\Delta t = 0$. This is the case, for instance, for the condensate fraction (critical exponent $\textit{v} = 1$) and for the specific heat capacity (critical exponent $\alpha = -1$). Indeed, according to (\ref{eta-ansatz}), the critical exponent actually determines not the power of the reduced temperature in front of the universal function, but rather the power of scaling of the deviation of the physical quantity from its thermodynamic-limit critical value in terms of the dispersion $\sigma^{(\infty)}$ of BEC fluctuations which can scale with anomalous dimension in the large-size limit, $\sigma^{(\infty)} \propto L^{\Delta_{\sigma}}$. The dispersion $\sigma^{(\infty)}$ depends on the temperature and size of the trap, as well as on its shape and boundary conditions, so that it uniquely incorporates effects of all various physical parameters of the system on the scale of critical fluctuations. 
         
\section{Conclusions}
The constraint-cutoff mechanism is the basic universal reason for the nonanalyticity and all other unusual critical phenomena of the BEC phase transition in an ideal gas. In terms of the Landau function \cite{Goldenfeld,Sinner}, i.e., the logarithm of the probability distribution of the order parameter, all these critical features originate from the constraint nonlinearity $\theta (N-\hat{n})$, i.e. due to the many-body Fock space cutoff in the canonical ensemble, which is responsible for the infinite potential wall in the effective fluctuation Hamiltonian and makes the Hamiltonian strongly asymmetric and nonanalytical even in the noninteracting gas. This constraint is directly related, through Noether's theorem, to the symmetry to be broken in the second-order phase transition and, hence, is the main reason for the phase transition and critical phenomena themselves. The nonanalyticity imposed by the constraint cutoff is crucially important. It cannot be taken into account perturbatively and makes the universal functions of physical quantities in the critical region so nontrivial that it is very difficult to find them directly, without explicit knowledge of the exact formulas which express the constraint-cutoff mechanism. It is worth stressing that the constraint-cutoff solution for the probability distribution (\ref{rhocut}) satisfies exactly the well-known recursion relation and is the exact, rigorous solution for the BEC statistics in an ideal gas in the canonical ensemble. It is not an approximation or a model. 

    The probability distribution of the condensate occupation, the order parameter, and all moments and cumulants of the BEC statistics as well as the thermodynamic quantities in the canonical ideal gas for any finite-size mesoscopic system of atoms in the trap can be scaled to the appropriate regular nontrivial critical functions which resolve the structure of the BEC phase transition in the critical region and converge fast to the corresponding universal functions in the thermodynamic limit (see also \cite{KKD-RQE}). We find exact analytical formulas for these universal functions in the whole critical region and their simple explicit approximations via confluent hypergeometric and parabolic cylinder functions in the central part of the critical region. That universality is described by the universal probability distribution (\ref{rhouniv}) of the total noncondensate occupation and exists for an ideal gas in any trap and for any boundary conditions. The particular shapes of the universal functions depend on a trap energy spectrum. The standard mean-field theory does not resolve the structure of the BEC phase transition at all. The described universality provides a complete and clear picture of the BEC statistics and thermodynamics that should not rely anymore only on the simulations for the finite-size systems with some particular values of numerous parameters. Hence, the long-standing problem of finding the universal structure of the critical region and, in particular, resolving the universal structure of specific heat of an ideal gas near the $\lambda$ point has a full analytical solution. 
     
     The self-similarity and constraint-cutoff mechanism of critical fluctuations discussed in the present paper are also generic for other second-order phase transitions and apply in interacting many-body systems. In an interacting gas, the shapes of the universal functions for the moments and cumulants of the BEC fluctuations and the thermodynamic quantities, in addition, depend on a deformation of the statistical distribution due to a feedback of the order parameter on the quasiparticle energy spectrum and correlations. We can describe these effects using nonperturbative-in-fluctuations theory based on the diagram technique and theorem on nonpolynomial averages in statistical physics proposed in \cite{KochLasPhys2007,KochJMO2007}. The self-similarity, universal functions, and the whole theory of critical phenomena in BEC statistics and thermodynamics outlined in this paper can be directly generalized to the case of BEC in a weakly interacting gas and other second-order phase transitions. That microscopic theory goes beyond the phenomenological renormalization-group theory and will be presented elsewhere. 
\ack
Support from the RFBI (grant 09-02-00909-a) and from the council on grants of the President of the Russian Federation for the support of the leading scientific schools of the Russian Federation (HIII-4485.2008.2) is greatly acknowledged.

\end{document}